\documentclass[10pt,conference]{IEEEtran}
\IEEEoverridecommandlockouts

\usepackage{cite}
\usepackage{amsmath,amssymb,amsfonts}
\usepackage{algorithmic}
\usepackage{graphicx}
\usepackage{textcomp}
\usepackage{xcolor}
\usepackage{acronym}
\usepackage{bm} 
\usepackage{url}
\usepackage{balance}
\usepackage{subfig}
\usepackage{tabularx}
\usepackage{booktabs,multirow} 
\usepackage{amssymb}
\usepackage{placeins}
\usepackage{pifont}
\usepackage{float}
\usepackage{multirow}
\usepackage{hyperref}
\usepackage{caption}
\usepackage{tikz}
\usepackage{graphicx}
\usetikzlibrary{shapes.geometric, arrows.meta, positioning,decorations.pathreplacing, positioning}
\usetikzlibrary{patterns}
\newcommand{\rayon}{2}

\acrodef{HRTF}{Head-Related Transfer Function}
\acrodef{HRIR}{Head-Related Impulse Response}
\acrodef{TrF}{Transfer Function}
\acrodef{TSE}{Target Speaker Extraction}
\acrodef{TF}{Time-Frequency}
\acrodef{RIR}{Room Impulse Responses}
\acrodef{STFT}{Short-Time Fourier transform}
\acrodef{ReLU}{Rectified Linear Unit}
\acrodef{DOA}{Direction of Arrival}
\acrodef{SI-SDR}{Scale Invariant Signal to Distortion Ratio }
\acrodef{FFT}{Fast Fourier Transform}
\acrodef{BSS}{Blind Source Separation}
\acrodef{MAE}{Mean Absolute Error}
\acrodef{WSJ}{Wall Street Journal}
\acrodef{SOFA}{Spatially Oriented Format for Acoustics}
\acrodef{SR}{Sampling Rate}
\acrodef{IR}{Impulse Response}
\acrodef{PESQ}{Perceptual Evaluation of Speech Quality}
\acrodef{MOS}{Mean Opinion Score}
\acrodef{ILD}{Interaural Level Difference}
\acrodef{ITD}{Interaural Time Difference}
\acrodef{STOI}{Short-time Objective Intelligibility}
\acrodef{RTF}{Relative Transfer Function}
\acrodef{DNN}{Deep Neural Network}
\acrodef{SIR}{Signal-to-Interference Ratio}
\acrodef{p.d.f.}{Probability Density Function}
\acrodef{LCMV}{Linearly Constrained Minimum Variance}
\acrodef{BLCMV}{Binaural Linearly Constrained Minimum Variance}
\acrodef{RIR}{Room Impulse Response}
\acrodef{LS}{LibriSpeech}
\acrodef{MLS}{Multilingual LibriSpeech}
\acrodef{MIMO}{Multi input Multi output}
\acrodef{Bi-TSE}{Binaural Target Speaker Extraction}
\acrodef{Bi-TSE-HRTF}{Binaural Target Speaker Extraction using HRTFs}
\acrodef{RI}{Real-Imaginary}
\acrodef{MOS}{Mean Opinion Score}

\def\BibTeX{{\rm B\kern-.05em{\sc i\kern-.025em b}\kern-.08em
    T\kern-.1667em\lower.7ex\hbox{E}\kern-.125emX}}
\begin{document}

\title{BINAURAL TARGET SPEAKER EXTRACTION USING INDIVIDUALIZED HRTF\thanks{This work was supported by the Israel Science Foundation (ISF) and the German Research Foundation (DFG) through the ISF–DFG Joint Research Program, Grant No.~1280/25.}}

\author{\IEEEauthorblockN{Yoav Ellinson}
\IEEEauthorblockA{\textit{Faculty of Engineering}, \textit{Bar-Ilan University}, Israel \\
yoav.ellinson@biu.ac.il; 0009-0001-6116-2869}
\and
\IEEEauthorblockN{Sharon Gannot}
\IEEEauthorblockA{\textit{Faculty of Engineering}
\textit{Bar-Ilan University}, Israel \\
sharon.gannot@biu.ac.il; 0000-0002-2885-170X}

}

\maketitle

\begin{abstract}
In this work, we address the problem of binaural target-speaker extraction in the presence of multiple simultaneous talkers. We propose a novel approach that leverages the individual listener’s \ac{HRTF} to isolate the target speaker. The proposed method is speaker-independent, as it does not rely on speaker embeddings.
We employ a fully complex-valued neural network that operates directly on the complex-valued \ac{STFT} of the mixed audio signals, and compare it to a \ac{RI}-based neural network, demonstrating the advantages of the former.
We first evaluate the method in an anechoic, noise-free scenario, achieving excellent extraction performance while preserving the binaural cues of the target signal. We then extend the evaluation to reverberant conditions. Our method proves robust, maintaining speech clarity and source directionality while simultaneously reducing reverberation.
A comparative analysis with existing binaural \ac{TSE} methods shows that the proposed approach achieves performance comparable to state-of-the-art techniques in terms of noise reduction and perceptual quality, while providing a clear advantage in preserving binaural cues.
\noindent Demo-page: \url{https://bi-ctse-hrtf.github.io}
\end{abstract}


\section{Introduction}
\label{sec:intro}
Humans have a remarkable ability to focus on the voice of a single speaker, even in complex auditory environments with multiple speakers. Such a scenario is often referred to as the cocktail party problem \cite{cherry1953}. While this ability may pose a challenge even to those with normal hearing, it presents a far greater obstacle for individuals with hearing impairments. \ac{Bi-TSE} aims to selectively extract the desired speaker from multiple competing speakers in a binaural setting.
In this context, preserving the binaural cues is critical: hearing aids and their associated algorithms must maintain the listener’s spatial awareness to ensure natural sound perception \cite{blauert1997spatial,binauralchapter}. 
Most hearing aids employ either a single microphone or multiple microphones per earpiece and can be broadly categorized into two groups:
1) Bilateral systems, where each device operates independently and provides a single-channel output per earpiece; 2) Shared systems, where all microphones across both earpieces are jointly utilized, yielding a dual-channel output.

Two important spatial cues are the \ac{ILD} and \ac{ITD}, which capture magnitude and arrival-time differences between the left and right ears. In bilateral systems, in which each device processes its own channel independently, it is particularly challenging to preserve binaural cues. In contrast, \ac{MIMO} solutions jointly process both channels, yielding more natural and spatially accurate outputs. For example, \cite{7372431} presents a \ac{BLCMV} beamformer that separates sources while maintaining binaural cues. More recent work, such as \cite{han2020real}, employs a \ac{DNN} to process all microphone signals and produce left- and right-channel outputs, effectively preserving spatial information.

Unlike \ac{BSS} algorithms, which separate all sources, \ac{TSE} algorithms leverage prior knowledge on the target speaker(s) to isolate their voice. This prior knowledge—commonly referred to as a “clue”~\cite{10113382}—can take various forms. A widely used clue is a pre-recorded speech sample, often referred to as an enrollment utterance, which provides speaker-specific features~\cite{8462661,wang2018deep,9909545,9067003}. Clues may also come from other modalities, such as visual data. For example, \cite{chung2020facefilter} uses a face image to extract the corresponding speech, while \cite{ochiai2019multimodal} combines audio and visual clues, with the audio clue being an enrollment utterance and the visual clue a video of the speaker speaking within the mixture.

While examining prior works, such as \cite{9054683}, we observed a particular challenge in separating two voices with highly similar characteristics, for example, speakers of the same gender. Consequently, we opted for an approach that does not rely on speaker–embedding–based clues. Extending beyond a single microphone enables us to exploit spatial information and more effectively isolate the desired speaker. For instance, \cite{eisenberg2025end} uses either the \ac{RTF} or the \ac{DOA} of the target speaker relative to a microphone array, while \cite{wang2025leveraging,mengbinaural} rely on the \ac{DOA} of the desired speaker and use it as a spatial clue. These spatially guided methods consistently outperform single-microphone solutions.

In this work, we propose \acf{Bi-TSE-HRTF}, a target speaker extraction method that exploits the individual listener’s \ac{HRTF}, aligned with the target speaker’s \ac{DOA}, as the primary extraction clue. Our approach employs a complex‑valued \ac{DNN} that operates directly in the \ac{STFT} domain. 
The proposed architecture produces binaural audio outputs, ensuring that spatial cues are fully preserved. Moreover, in reverberant environments, the model isolates the direct path's \ac{HRTF}, thereby dereverberating the signal and hence further enhancing its binaural characteristics. The method is thoroughly assessed using simulated \acp{RIR} and standard \ac{HRTF} databases, and compared with two competing methods, a beamformer-based solution and a \ac{DNN}-based solution.

\section{Problem Formulation}
\label{Problem Formulation }

We address the problem of multiple concurrent speakers captured by two microphones mounted on the listener’s ears, assuming a noiseless environment and a fixed source–listener radius. For clarity, we focus on the case of two concurrent speakers, as formulated below.

Denote the desired speaker’s \ac{DOA} by $(\theta_d,\phi_d)$ and the interferer’s by $(\theta_i,\phi_i)$. Let $\bm{x}_b(n,k)$ denote the \ac{STFT} of the binaural signal, with $n\in[0,N-1]$ and $k\in[0,K-1]$ the time-frame and frequency-bin indices, respectively. The binaural signal is therefore formulated as:
\begin{equation}
\bm{x}_b(n,k) \;=\;  s_d(n,k)\,\bm{h}(\theta_d,\phi_d,k)+s_i(n,k)\,\bm{h}(\theta_i,\phi_i,k)
\label{eq:1}
\end{equation}  
where $s_d(n,k)$ and $s_i(n,k)$ are the anechoic speech signals of the desired speaker and the interfering speaker, respectively. 
Let $\gamma \in \{d,i\}$ denote the speaker index, where $d$ refers to the desired source and $i$ to the interferer. The acoustic transfer from source $\gamma$ to both earpieces is described by $\bm{h}(\theta_{\gamma},\phi_{\gamma},k)$, which models the \ac{TrF} as a superposition of \acp{HRTF} corresponding to the \acp{DOA} of all acoustic reflections in the enclosure, with the \ac{HRTF}, $\bm{h}_{\text{hrtf}}(\theta,\phi,k)$, characterizeing the listener’s response to a wave arriving from \ac{DOA} $(\theta,\phi)$.

According to the precedence effect \cite{haas1951influence}, the direct-path \ac{HRTF} dictates the perceived source \ac{DOA}. Thus, although $\bm{h}(\theta_{\gamma},\phi_{\gamma},k)$ encompasses all reflections, it is indexed by the direct-path \ac{DOA} for clarity.
Finally, define $\bar{\bm{h}}(\theta_{\gamma},\phi_{\gamma},t)$, with $t$ the filter's tap index, as the time-domain representation of $\bm{h}(\theta_{\gamma},\phi_{\gamma},k)$, and similarily define the \ac{HRIR} at the \ac{DOA} $\theta,\phi$ as  $\bar{\bm{h}}_{\text{hrir}}(\theta,\phi,t)$.
Finally, the target signal, $\tilde{\bm{s}}_d=[\tilde{s}_d^{l}, \tilde{s}_d^{r}]^\top$, is a concatenation of the desired speech signal as received at the left and right ears, respectively.

Given the individual \ac{HRTF} of the hearing-aid wearer corresponding to the desired source \ac{DOA} $(\theta_d,\phi_d)$, denoted by the binaural response $\bm{h}_{\text{hrtf}}(\theta_d,\phi_d,k)$, our method aims to extract $\tilde{\bm{s}}_d$ from the mixed signal $\bm{x}_b$.
The extracted signal is denoted by $\hat{\tilde{\bm{s}}}_d$. It is required that the binaural output satisfy the following conditions: 
1) preservation of the spatial cues of the desired source, and 
2) dereverberation of the desired speech output by retaining only the direct-path \ac{HRTF}.
In this work, we focus on \emph{individualized} \ac{HRTF} database, namely the set of \acp{HRTF} of the hearing aid wearer across all azimuth and elevation directions, which should be given to the algorithm. \ac{HRTF} ersonalization using anthropometric measurements \cite{zotkin2003hrtf}.

\section{Proposed Method}
\label{Proposed Method}
We propose a \ac{DNN}-based model to extract a desired speech signal corresponding to a given \ac{HRTF} clue.

\subsection{Architecture}
\label{Architecture}
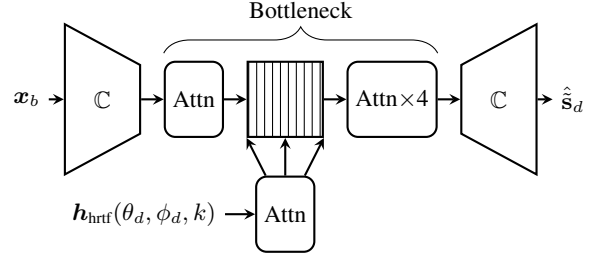
\begin{figure}[t!]
    \centering
    \begin{tikzpicture}[
    block/.style={draw, thick, minimum height=1cm,rounded corners, minimum width=.5cm},
    thin_block/.style={draw, thick, minimum height=1cm},
    arrow/.style={thick,->,>=stealth},
    sum/.style={draw, circle, inner sep=0pt, minimum size=1cm, node distance=2cm},
    dashedarrow/.style={thick, dashed,->,>=stealth},
    encoder/.style={draw, thick, trapezium,shape border rotate=270, trapezium stretches=true, minimum height=1cm, minimum width=2cm},
    decoder/.style={draw, thick, trapezium,shape border rotate=90, trapezium stretches=true, minimum height=1cm, minimum width=2cm}
]

\node[encoder] (enc) at (0,0) {\(\mathbb{C}\)};  
\node[thin_block,draw=none,left=.2cm of enc] (x) {$\bm{x}_b$}; 
\node[block, right=0.3cm of enc] (attn_mix) {Attn};
\node[thin_block, right=0.3cm of attn_mix,text=white,pattern=vertical lines,minimum width=1cm] (bn) {};
\node[block, below=0.5cm of bn,align=center] (attn_hrtf) {Attn};
\node[thin_block,draw=none,left=.4cm of attn_hrtf] (hrtf) {$\bm{h}_{\text{hrtf}}{(\theta_d,\phi_d,k)}$}; 
\node[block, right=0.3cm of bn] (post_attn) {Attn$\times$4};
\node[decoder, right=0.3cm of post_attn] (dec) {\(\mathbb{C}\)};
\node[thin_block,draw=none,right=.2cm of dec] (out) {$\hat{\tilde{\mathbf{s}}}_d$};

\draw[arrow] (x) -- (enc);
\draw[arrow] (enc) -- (attn_mix);
\draw[arrow] (attn_mix) -- (bn);
\draw[arrow] (bn) -- (post_attn);
\draw[arrow] (hrtf) -- (attn_hrtf);

\draw[arrow] (attn_hrtf)--(bn.south east);
\draw[arrow] (attn_hrtf)--(bn.south west);
\draw[arrow] (attn_hrtf)--(bn);

\draw[arrow] (post_attn) -- (dec);
\draw[arrow] (dec) -- (out);
\draw[decorate,decoration={brace,amplitude=10pt}] 
  ([yshift=.1cm]attn_mix.north west) -- ([yshift=.1cm]post_attn.north east) 
  node[midway,above=10pt] {Bottleneck};
\end{tikzpicture}
    \addtolength{\belowcaptionskip}{-14pt}
    \caption{A block diagram of the proposed method, where $\bm{h}_{\text{hrtf}}{(\theta_d,\phi_d,k)}$ denotes the \ac{HRTF} of the desired speaker’s DOA, $\bm{x}_b$ represents the mixed binaural signal, and  $\hat{\tilde{\bm{s}}}_d$ represents the estimated desired signal, both in the \ac{STFT} domain.}
    \label{fig:block_diagram}
\end{figure}

The proposed model architecture is built upon the approach presented in \cite{eisenberg2025end}, with modifications to handle dual-channel, complex-valued inputs rather than treating the real and imaginary components independently. The proposed model employs a U-Net architecture, incorporating the \ac{HRTF} at the bottleneck. The individual modules comprising the architecture are described in detail below and depicted in Fig.~\ref{fig:block_diagram}.


\vspace{1pt}\noindent\textbf{Encoder--Decoder Blocks of the U-Net:}
The encoder consists of five down-convolution blocks, each comprising a 2D convolutional layer, global layer normalization, and a complex-valued ReLU activation defined as
\vspace{-5pt}
\begin{equation}
\mathbb{C}\text{ReLU}(z)
= \text{ReLU}\!\left(\mathcal{R}(z)\right)
+ \mathrm{j}\,\text{ReLU}\!\left(\mathcal{I}(z)\right),
\end{equation}
where $\text{ReLU}(x)=\max(0,x)$, $\mathcal{R}(z)$ and $\mathcal{I}(z)$ denote the real and imaginary parts of $z$, and $\mathrm{j}$ is the imaginary unit.
The decoder mirrors the encoder with five up-convolution blocks based on transposed convolutions.
Skip connections are employed between corresponding encoder and decoder blocks to improve reconstruction accuracy.
The decoder further includes a final convolutional layer without an activation function, allowing the output to represent both positive and negative real and imaginary values, which is required for the complex \ac{STFT} representation.

\vspace{1pt}\noindent\textbf{Bottelneck:}
The extraction clue is applied at the bottleneck of the encoded input features, with self-attention layers used both before and after its application. Unlike conventional approaches, these self-attention layers operate directly on complex-valued tensors by replacing the softmax activation with the complex-valued $\mathbb{C}\text{ReLU}$.
To capture cross-channel (left–right) information in the \ac{HRTF}, a self-attention mechanism is employed, enabling the channels to attend to one another via learned weights. The resulting matrix is flattened and projected via a fully-connected layer to match the dimensionality of the encoded mixture, yielding the \ac{HRTF} embedding. Similarly, self-attention is applied to the encoded mixed signal. The replicated \ac{HRTF} embedding is then combined with each time frame of the encoded mixture through element-wise multiplication, thereby imposing the extraction clue. The resulting tensor is processed by four additional self-attention layers and a fully connected layer before decoding.

\subsection{Loss function}
The model was initially trained using only the \ac{SI-SDR} loss~\eqref{eq:sisdr}.
\vspace{-5pt}
\begin{equation}
    \text{SI-SDR}(\bm{s}, \hat{\bm{s}}) = 20 \log_{10}\left(\frac{\| \beta \bm{s} \|}{\| \beta \bm{s} - \hat{\bm{s}} \|}\right);   \quad \beta = \frac{\hat{\bm{s}}^\top \bm{s}}{\|\bm{s}\|^2}\,,
    \label{eq:sisdr}
\end{equation}
where $\bm{s}$ and $\hat{\bm{s}}$ are vectors comprising all time-domain samples of the target and estimated signals, respectively.
However, relying solely on the \ac{SI-SDR} loss proved challenging in preserving accurate spectral details, resulting in noticeable musical noise artifacts in the extracted signals. Musical noise refers to tonal distortions that occur due to inaccuracies or rapid variations in the estimated signal's frequency-domain representation \cite{cappe1994elimination}. Such artifacts are commonly observed in signal enhancement tasks when spectral continuity and consistency are not adequately enforced. To mitigate these artifacts, we introduced an additional frequency-domain loss, the $L1$-\ac{MAE} loss:
\vspace{-5pt}
\begin{equation}
    \mathcal{L}_{\text{MAE}} = \frac{1}{2KN} \sum_{c=\{l,r\}}\sum_{k=0}^{K-1}\sum_{n=0}^{N-1}\left| |\tilde{s}_d^c(n,k)| - |\hat{\tilde{s}}_d^c(n,k)| \right|.
    \label{eq:l1}
\end{equation}
Here, $c=\{l,r\}$ denotes the output channels, left or right.
To improve data utilization, both speakers were extracted from each mixed signal in each training step by alternating the \ac{HRTF} clues and averaging the \ac{SI-SDR} losses.
The overall training loss is therefore defined in~\eqref{eq:loss}, as a weighted combination of the \ac{SI-SDR} and $L1$-MAE losses:
\begin{equation}
\mathcal{L} = \frac{1}{2}\sum_{\gamma=\{d,i\}}\left( -\text{SI-SDR}(\tilde{\bm{s}}_{\gamma}, \hat{\tilde{\bm{s}}}_{\gamma}) + \alpha \cdot \mathcal{L}_{\text{MAE}}(\tilde{\bm{s}}_{\gamma}, \hat{\tilde{\bm{s}}}_{\gamma})\right).
\label{eq:loss}
\end{equation}
Due to the complexity of multi-objective optimization in high-dimensional space, we adopted a multi-stage training strategy, in which we first train the algorithm using only the \ac{SI-SDR} loss (i.e., $\alpha = 0$). Once a sufficient performance level has been reached, the $L_1$-MAE loss is introduced (namely, $\alpha > 0$), improving both spectral accuracy and audio quality. The final stage fine-tunes the model using only the \ac{SI-SDR} loss, as the $L_1$-MAE loss is prone to overfitting to the training data.

\section{Experimental Study}
\label{Experimental study}
In this section, we describe the data-generation process and experimental setup, evaluate the performance of the proposed approach, and compare it with previous \ac{TSE} methods.

\subsection{Data Generation}
\label{Signals generation}
\vspace{1pt}\noindent\textbf{Binaural Impulse Rsponses:}
The binaural \acp{IR} are formed by superimposing attenuated and time-shifted \acp{HRIR}, corresponding to acoustic reflections generated via the image method~\cite{allen1979image}, as expressed in~\eqref{eq:hrtf_rir}.
%
\vspace{-5pt}
\begin{equation}
    \bar{\bm{h}}(\theta_{\gamma},\phi_{\gamma},t) = \sum_{\ell=0}^{L}a_\ell\cdot\bar{\bm{h}}_{\text{hrir}}(\theta_\ell,\phi_\ell,t-\tau_\ell)
    \label{eq:hrtf_rir}
\end{equation}
where $L$ is the reflection order.
The direct path ($\ell=0$) corresponds to the \ac{HRIR} of the desired or interfering source at the listener’s head for the \acp{DOA} $(\theta_{\gamma}, \phi_{\gamma})$, and the gain $a_\ell$ and the time delay $\tau_\ell$, obtained from the image method, are applied by scaling and time-shifting each HRTF component. 

\vspace{1pt}
\noindent\textbf{The RIEC HRTF Dataset \cite{Kanji,RIECwebpage}:}
We utilized the RIEC database, which comprises measurements from 105 subjects, each with 865 directions. For this study, we focused on a single subject. The \ac{SOFA} \ac{HRTF} Toolbox~\cite{majdak2013spatially,majdak2022spatially} facilitates easy retrieval of the appropriate \ac{HRTF} for any desired \ac{DOA}.
We employed the SofaMyRoom framework \cite{barumerli2021sofamyroom} to simulate both anechoic and reverberant conditions, following the formulation in \eqref{eq:hrtf_rir}. This framework positions a listener, characterized by their measured \ac{HRTF}, inside a shoebox-shaped room with configurable wall materials to achieve a specified reverberation time. It also supports zero-order simulations for anechoic scenarios. The reverberation time was sampled as $T_{60} \sim \mathcal{U}[0.2,0.8]$, and the listener position is drawn uniformly within the room. The overall simulation setup is illustrated in Fig.~\ref{fig:mesurment_system}.

\vspace{1pt}\noindent\textbf{Speech Datasets}
We trained and validated our model on the WSJ0‑CSR1 dataset \cite{garofolo2007csr}.
\begin{figure}[b]
    \vspace{-19pt}
    \centering
    \input{tikz_files/protractor}
    \caption{A simulation illustration: two concurrent speakers: the desired speaker at \( \theta_d = 40^\circ \) (left) and the interferer at \( 
    \theta_i = -30^\circ \) (right), both at elevation \( \phi \). Images source: \href{https://www.freepik.com}{freepik.com}}
    \label{fig:mesurment_system}
\end{figure}

\subsection{Algorithm Setup}
The speech signals, sampled at 16 kHz, are randomly selected, cropped, or zero-padded to 5~s, then transformed with a 512-sample \ac{STFT} (75\% overlap).
The model is trained in a supervised manner: in the anechoic setting, targets are generated by convolving the desired speech with the corresponding \ac{HRTF}; in the reverberant setting, by simulating a room and keeping only the first arrival (zero-order reflection).
The \acp{HRTF}, used as the extraction clue, are downsampled from 48~kHz to 16~kHz and converted to the frequency domain via \ac{FFT}.

\begin{figure*}[ht!]
    \centering
    \subfloat[Reverberant Mixture]{\includegraphics[width=0.2\linewidth,trim= 0 10 70 45,clip]{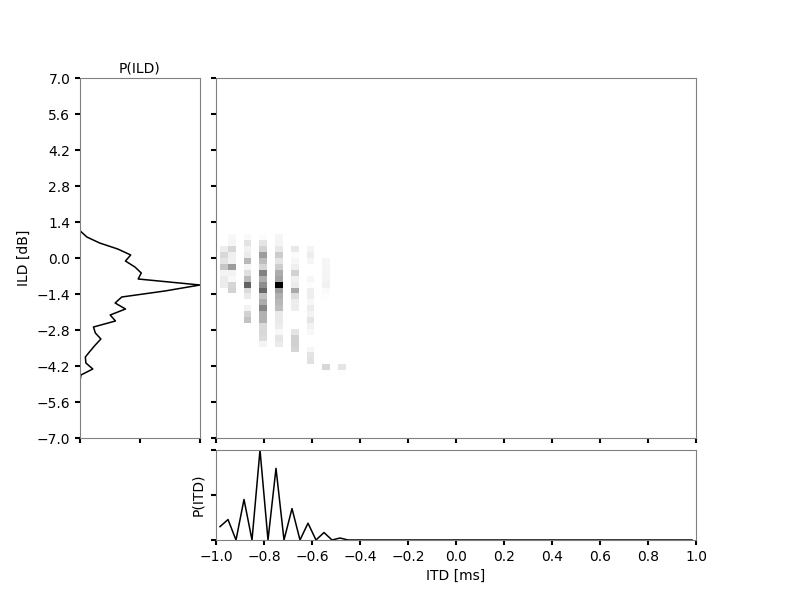}}
    \subfloat[$\tilde{\bm{s}}_d$]{\includegraphics[width=0.2\linewidth, trim= 0 10 70 45,clip]{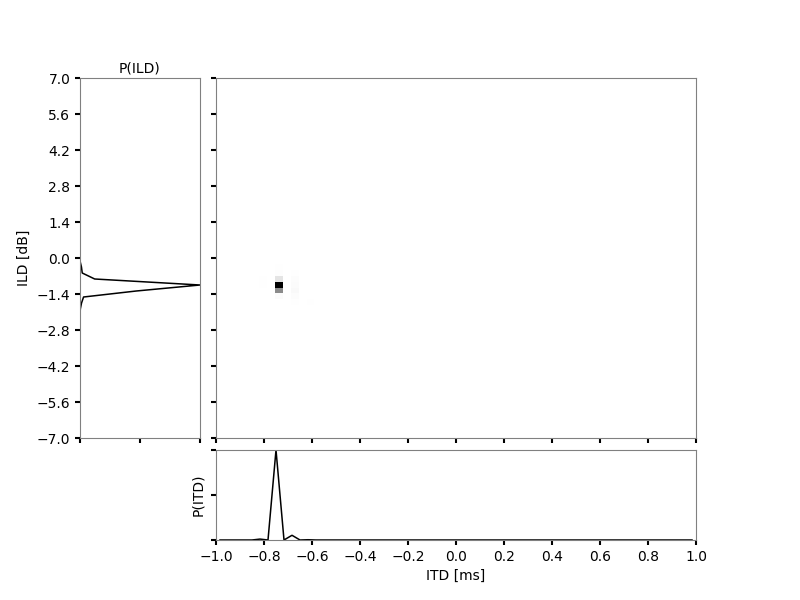}} 
    \subfloat[$\hat{\tilde{\bm{s}}}_d$]{\includegraphics[width=0.2\linewidth, trim= 0 10 70 45,clip]{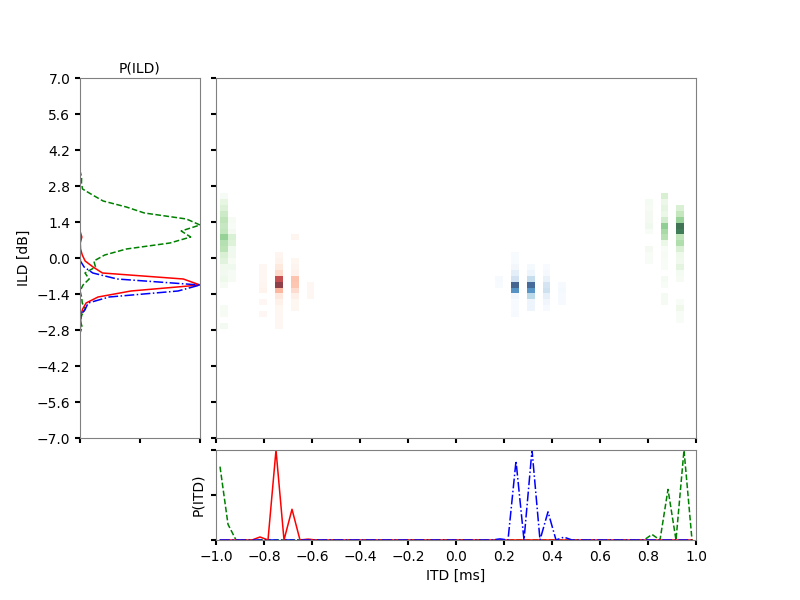}}
    \subfloat[$\tilde{\bm{s}}_i$]{\includegraphics[width=0.2\linewidth, trim= 0 10 70 45,clip]{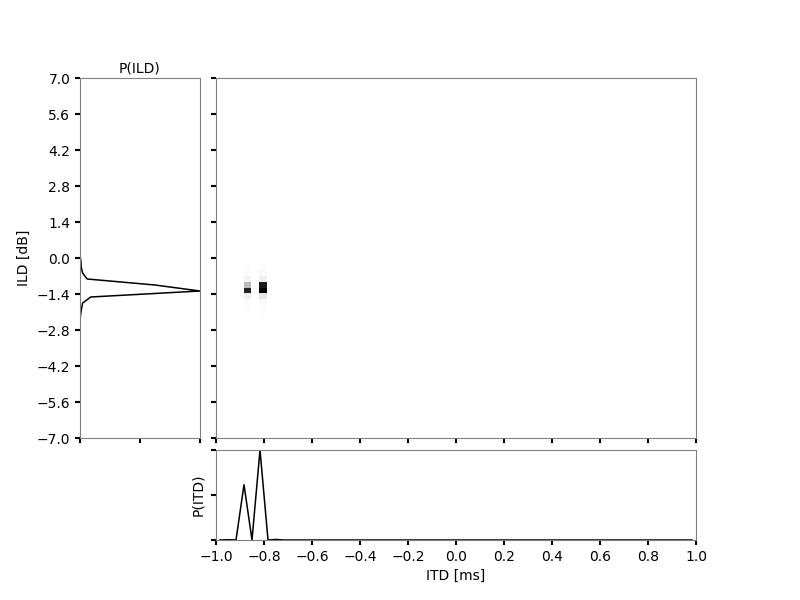}}
    \subfloat[$\hat{\tilde{\bm{s}}}_i$]{\includegraphics[width=0.2\linewidth, trim= 0 10 70 45,clip]{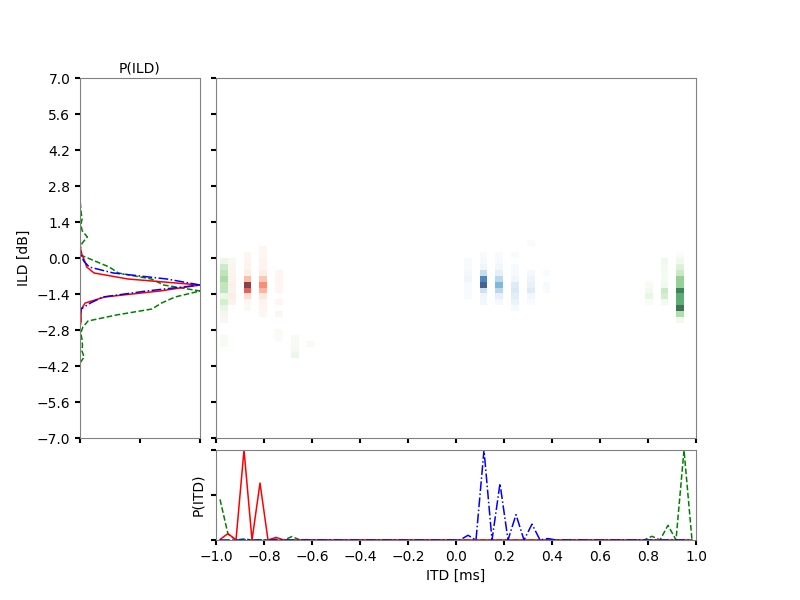}}
    \vspace{-0.1cm}
 
    \begin{tikzpicture}[baseline]
    \draw[black, thick] (0,0.15) -- (0.6,0.15);
    \node[right=3mm] at (0.6,0.15) {\small{Ground-truth}};

    \draw[red, thick] (3.4,0.15) -- (4.0,0.15);
    \node[right=3mm] at (4.0,0.15) {\small{\ac{Bi-TSE-HRTF} (proposed)}};

    \draw[blue, thick, dash dot] (8,0.15) -- (8.6,0.15);
    \node[right=3mm] at (8.6,0.15) {\small{BDE-BiTSE\cite{wang2025leveraging}}};

    \draw[thick, dashed, color={rgb,255:red,11; green,162; blue,52}] (12,0.15) -- (12.6,0.15);
    \node[right=3mm] at (12.6,0.15) {\small{BLCMV Beamformer \cite{7372431}}};
\end{tikzpicture}

    \addtolength{\belowcaptionskip}{-6pt}
    \addtolength{\abovecaptionskip}{-7pt}
    \caption{The joint \acp{p.d.f.} of ITD [ms] and ILD [dB] for $\theta_d = -60^\circ$, $\theta_i = -90^\circ$, and $\phi_{d,i} = -10^\circ$, with $T_{60}=0.63$~s, for the frequency band centerd at 500~Hz, where the ILD is less pronounced. $\tilde{\bm{s}}_d$ and $\tilde{\bm{s}}_i$ denotes the anechoic target signals. Graphs produced by~\cite{faller2004source}.}
    \label{fig:binaural_cues}
\end{figure*}
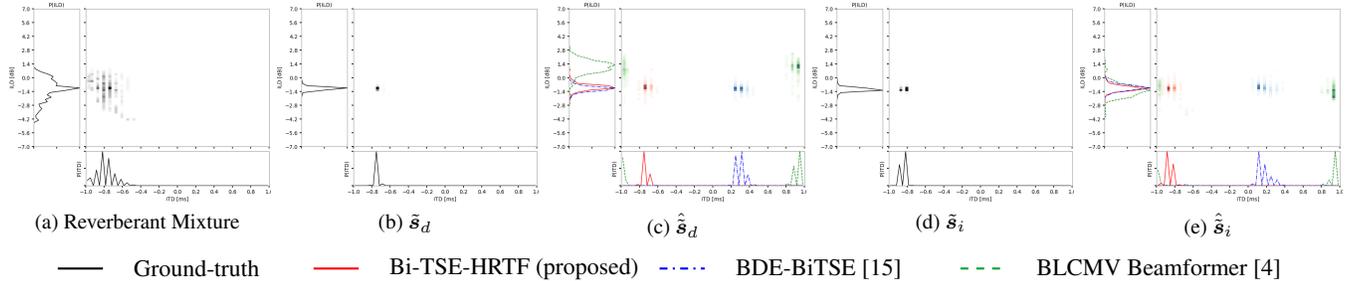

\subsection{Metrics}
\label{sec:metrics}
To evaluate our method, we use \ac{SI-SDR}~\cite{le2019sdr}, wide- and narrow-band \ac{PESQ}~\cite{rix2001perceptual}, and \ac{STOI}~\cite{taal2011algorithm} as intrusive metrics of speech quality and intelligibility.
In addition, we employ DNS-MOS~\cite{reddy2021dnsmos}, a non-intrusive \ac{DNN}-based \ac{MOS} predictor estimating speech quality (SIG), background noise quality (BAK), and overall quality (OVRL).
Furthermore, to assess binaural cue preservation, we compare the \ac{ILD}–\ac{ITD} \acp{p.d.f.} of the extracted and target signals, where close agreement indicates preserved spatial cues. These metrics are computed using the procedure in \cite{faller2004source}.

\begin{table*}[htbp]
    \centering
    \caption{Each evaluation dataset contains 1500 speech mixtures, with both speakers extracted, yielding 3000 samples. Mixtures were created with \ac{SIR} drawn from $\mathcal{U}[-5,5]$ dB, and 100\% overlap. Best scores are highlighted in boldface.}
    \vspace{-4pt}
    \scriptsize
    \setlength{\tabcolsep}{2pt}
    \renewcommand{\arraystretch}{0.85}
    \resizebox{0.99\textwidth}{!}{
    \begin{tabular}{c l c c c c c c c}
        \toprule
        \textbf{Dataset} & \textbf{Method}  & \textbf{DNSMOS (OVRL\textbackslash SIG\textbackslash BAK)} & \textbf{SI-SDR (dB)} $\uparrow$ & \textbf{WB PESQ} $\uparrow$ & \textbf{NB PESQ} $\uparrow$ & \textbf{STOI} $\uparrow$ & \textbf{$\Delta$ILD [dB]} $\downarrow$ & \textbf{$\Delta$ITD [mS]} $\downarrow$ \\
        \midrule



        \multirow{6}{*}{WSJ0}
            & Unprocessed & (2.845\textbackslash 3.376\textbackslash 3.411) & 0 & 1.25 & 1.57 & 0.72 & - & - \\
            & BLCMV Beamformer & (3.044\textbackslash 3.444\textbackslash 3.748) & -0.42 & 3.48 & 3.84 & 0.95 & 0.745 & 0.613 \\
            & BDE-BiTSE & \textbf{(3.230\textbackslash 3.543\textbackslash 3.951}) & \textbf{25.59} & \textbf{3.77} & \textbf{4.01} & \textbf{1.0} & 0.126 & 0.484 \\
            & Bi-TSE-DOA & (2.965\textbackslash 3.376\textbackslash 3.660) & 16.44 & 2.43 & 3.09 & 0.97 & 0.165 & 0.990 \\
            & Bi-TSE-HRTF-RI & (3.110\textbackslash 3.504\textbackslash 3.764) & 21.39 & 2.93 & 3.57 & 0.99 & 0.085 & 0.0044 \\
            & \ac{Bi-TSE-HRTF} (proposed) & (3.18\textbackslash 3.501\textbackslash 3.911) & 21.41 & 3.47 & 3.88 & 0.99 & \textbf{0.0756} & \textbf{0.0035} \\
        \midrule

        \multirow{6}{*}{WSJ0 (Reverb)}
            & Unprocessed & (2.348\textbackslash 3.069\textbackslash 2.783) & -1.9 & 1.15 & 1.44 & 0.69 & - & - \\
            & BLCMV Beamformer & (1.840\textbackslash 2.612\textbackslash 2.038) & -1.11 & 1.35 & 1.88 & 0.74 & 0.430 & 0.593 \\
            & BDE-BiTSE & \textbf{(3.141\textbackslash 3.438\textbackslash 3.952)} & \textbf{14.8} & \textbf{3.09} & \textbf{3.56} & \textbf{0.98} & 0.204 & 0.386 \\
            & Bi-TSE-DOA & (2.681\textbackslash 3.196\textbackslash 3.324) & 8.29 & 1.82 & 2.44 & 0.91 & 0.296 & 0.990 \\
            & Bi-TSE-HRTF-RI & (3.030\textbackslash 3.362\textbackslash 3.838) & 11.18 & 2.01 & 2.74 & 0.95 & \textbf{0.174} & 1.001 \\
            & \ac{Bi-TSE-HRTF} (proposed) & (2.86\textbackslash 3.348\textbackslash 3.481) & 10.67 & 2.17 & 2.79 & 0.94 & 0.186 & \textbf{0.0082} \\
        \bottomrule
    \end{tabular}
    }
    \label{table:results}
    \vspace{-14pt}
\end{table*}

\begin{table}[htp]
\centering
\addtolength{\belowcaptionskip}{-2pt}
\addtolength{\abovecaptionskip}{-1pt}
\caption{Comparison of $\Delta$ITD, $\hat{\theta}$, and $\Delta\theta$ for $\theta_d = -60^\circ$.}
\begin{tabular}{lccc}
\toprule
\textbf{Method} & $\Delta$ITD [ms] & $\hat{\theta}$ [°] & $\Delta\theta$ [°] \\
\midrule
BLCMV Beamformer & 1.7 & $>90$ & $>150$ \\
BDE-BiTSE & 1.0667 & 30 & 90\\
\ac{Bi-TSE-HRTF}& \textbf{0.0} & \textbf{-60} & \textbf{0} \\
\bottomrule
\vspace{-10mm}
\label{tab:delta_thata}
\end{tabular}
\end{table}

\subsection{Competing Methods}
To ensure a fair comparison, all baseline methods were evaluated under an identical setup, namely binaural \ac{TSE} with only two microphones (one per ear). Accordingly, we compared our method against the \ac{BLCMV} beamformer~\cite{7372431} and a recent \ac{DNN}-based model~\cite{wang2025leveraging}, both specifically designed for this scenario. The competing \ac{DNN}, referred to as BDE-BiTSE, was trained following the procedure described in the paper, using the exact same training data as ours. We ensured that the strongest variant reported in the paper (BDE+CDF+IPD+SDF) was used for comparison.

As an ablation study, we trained two variants of our model. The first, Bi-TSE-HRTF-RI, is a real-valued model in which the complex-valued input features are decomposed into their \ac{RI} components and concatenated, yielding four input channels rather than two. The same decomposition is applied to the \ac{HRTF} clue. The second, Bi-TSE-DOA, is a complex-valued model that incorporates the \ac{DOA} as a clue rather than the \ac{HRTF}. Here, the \ac{DOA} is represented as a one-hot-encoded vector with the same resolution as the available \acp{HRTF} and passed through a fully connected layer to match the bottleneck's dimensionality. For both variants, the loss function~\eqref{eq:loss} remains unaltered, with the target defined as the binaural rendering of the desired speech signal using the correct \ac{HRTF}.

\subsection{Results}
\label{Results}
In Table~\ref{table:results}, we present a comprehensive comparison of our method with competing algorithms in both anechoic and reverberant conditions. In terms of \ac{SI-SDR}, \ac{PESQ}, and DNS-MOS, our approach achieves competitive results in the anechoic setting, but underperforms in reverberant environments compared with BDE-BiTSE. Nevertheless, our model clearly preserves binaural cues, outperforming the baseline methods, as evidenced by the \ac{ITD} and \ac{ILD} measures.
The importance of minimizing $\Delta$\ac{ITD} is evident in systems that do not explicitly constrain it. Since \ac{ITD} reflects interaural arrival-time differences~\cite{binauralchapter}, inaccuracies lead to spatial errors and shifts in the perceived target location. To quantify this effect, we estimate the output \ac{DOA} by computing \ac{ITD} values for all candidate directions, forming a \ac{DOA}–\ac{ITD} lookup table, and comparing the estimated direction $\hat{\theta}$ to the ground truth. Owing to the computational cost, we report a single illustrative example in Table~\ref{tab:delta_thata}. The results indicate that competing methods exhibit substantial localization errors, whereas the proposed approach preserves spatial consistency.

Figure~\ref{fig:binaural_cues} shows the joint \ac{ITD}–\ac{ILD} \acp{p.d.f.} for the extracted and target signals. The close alignment of the proposed method with the target distribution confirms effective preservation of binaural cues. In contrast, competing methods exhibit noticeable deviations, resulting in shifted perceived source locations. The increased concentration of the joint distribution further indicates a dereverberation effect, with the direct path dominating the output.
\begin{figure}[hpt]
    \vspace{-12pt}
    \centering
    \addtolength{\belowcaptionskip}{-12pt}
    \addtolength{\abovecaptionskip}{-4pt}
    \includegraphics[width=0.8\linewidth,trim= 0 52 0 47,clip]{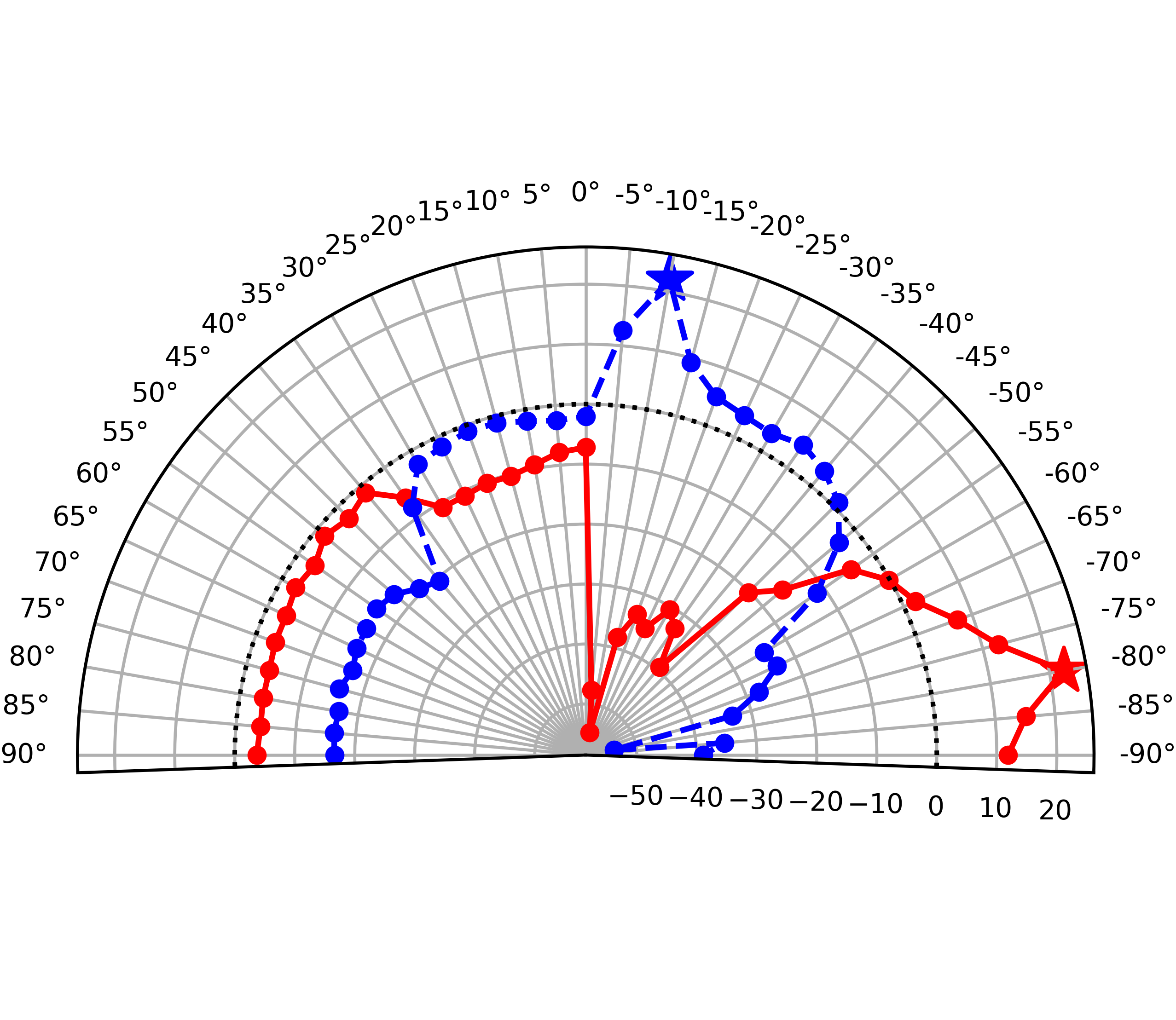}
    \caption{A polar plot of SI-SDR [dB] versus DOA, with the target at $\theta_d=-75^\circ$ (red) and the interferer at $\theta_i=-10^\circ$ (dashed blue). Stars mark true DOAs; }
    \label{fig:spatial_plots}
\end{figure}

Additionally, we evaluate two model variants: Bi-TSE-DOA and Bi-TSE-HRTF-RI.
The DOA-based model yields lower SI-SDR and PESQ, fails to suppress the interferer, and does not preserve binaural cues—especially under reverberation—highlighting the central role of \acp{HRTF} in our formulation.
The \ac{RI} model achieves SI-SDR and binaural cue preservation comparable to those of the complex-valued model in anechoic conditions, but yields lower \ac{PESQ} scores and degraded cue preservation under reverberation.
To demonstrate spatial resolution, we employ a ``scanning'' procedure in which signals are extracted using \acp{HRTF} over azimuths in $[-90^\circ,90^\circ]$, and the corresponding \ac{SI-SDR} is computed with respect to the target. Plotting these values yields a beam-pattern-like diagram (Fig.~\ref{fig:spatial_plots}) with a peak at the true \ac{DOA}, where the main-lobe width reflects the model’s stability to mild \ac{HRTF} deviations.

\section{Conclusions}
\label{Conclusions}
This paper presents a speaker-independent \ac{TSE} method that exploits the listener’s \acp{HRTF} in both anechoic and reverberant environments, providing a personalized solution. By conditioning the extraction process on individualized \acp{HRTF}, the proposed approach preserves binaural structure while maintaining speech quality. Comparative analyses with existing binaural \ac{TSE} methods demonstrate substantial improvements in spatial consistency, while preserving high speech quality. These results highlight the critical role of the listener’s \ac{HRTF} in anchoring spatial cues, enabling accurate and consistent target-speaker extraction across diverse acoustic conditions.

\bibliographystyle{IEEEbib}
\bibliography{refs25}
\end{document}